\documentclass[conference]{IEEEtran}
\usepackage{amsfonts}
\usepackage{cite}

\ifCLASSINFOpdf
  \usepackage[pdftex]{graphicx}
\else
\fi
\usepackage[cmex10]{amsmath}
\usepackage{amssymb}

\usepackage[tight,footnotesize]{subfigure}
\usepackage{fixltx2e}
\usepackage{multirow}

\makeatletter

\newcommand{\Rmnum}[1]{\expandafter\@slowromancap\romannumeral #1@}
\makeatother

\begin{document}
\title{Parallel MRI Reconstruction by Convex Optimization}

\author{\IEEEauthorblockN{Cishen Zhang and Ifat-Al Baqee}
\IEEEauthorblockA{Faculty of Engineering and Industrial Sciences,
Swinburne University of Technology\\
Hawthorn, VIC 3122, Australia\\
Email: cishenzhang@swin.edu.au and ibaqee@swin.edu.au}}
\maketitle

\begin{abstract}
In parallel magnetic resonance imaging (pMRI), to find a joint solution for the image and coil sensitivity functions is a nonlinear and nonconvex problem. A class of algorithms reconstruct sensitivity encoded images of the coils first followed by the magnitude only image reconstruction, e.g. GRAPPA. It is shown in this paper that, if only the magnitude image is reconstructed, there exists a convex solution space for the magnitude image and sensitivity encoded images. This solution space enables formulation of a regularized convex optimization problem and leads to a globally optimal and unique solution for the magnitude image reconstruction. Its applications to in-vivo MRI data sets result in superior reconstruction performance compared with other algorithms.
\end{abstract}

%

\IEEEpeerreviewmaketitle

\section{Introduction}
\IEEEPARstart{M}{agnetic} resonance imaging (MRI) is an advanced modality for obtrusive medical diagnosis
which provides very safe scanning, high spatial resolution and flexible contrast for displaying body proton mass.
To reduce the duration for scanning without compromising the image quality
has been an important and challenging problem in the MRI research. One approach is to implement multiple receiver coils
to accelerate MRI scans by acquiring simultaneously undersampled $k$-space data which is known as parallel MRI (pMRI).

The pMRI reconstruction using undersampled $k$-space data requires knowledge of spatial sensitivity functions of the multiple receiver coils, which are complex valued and determined by the coil instrumentation the imaged object \cite{roemer_1990}.
Several algorithms have been developed in past years for pMRI reconstruction,
one set of algorithms pre-estimate the complex valued sensitivity functions and use the estimated results to reconstruct the the complex valued image such as
SENSE \cite{pruessmann_1999} and its extensions.
Another set of algorithms estimate the sensitivity encoded images of each receiver coil first followed by a image reconstruction operation. Not only knowledge on sensitivity functions is implicitly prerequisite for these algorithms but also the reconstructed image function contains magnitude only values. Typical algorithms of this set are GRAPPA \cite{griswold_2002}
and its extensions using the sum-of-squares (SOS) operation \cite{roemer_1990}. Some recent algorithms of this group reconstruct the sensitivity encoded images by regularized optimization, e.g. \cite{lustig_2010, 
park_2012}. The third set of algorithms formulate the pMRI reconstruction into regularized optimization problem and jointly estimate the image and sensitivity functions, e.g. \cite{uecker_2008,knoll_2012,she_2012}.
Because of the cross product terms of the image and sensitivity functions, the formulated optimization problem is nonlinear and nonconvex and can only yield local solutions.

In our previous work \cite{cishen_2013}, it has been shown that, if only the magnitude of the image is reconstructed, the pMRI reconstruction can be formulated into a two-step convex optimization problem which resulted a global optimal solution. To ease the computational burdens of two step formulation, in this paper we present a single step iterative optimization formula for the problem which can lead to efficient computing of the solution as well as yield global optimal solution, also can outperform other pMRI reconstruction algorithms. Without loss of popularity and as the second group of algorithms including GRAPPA and its extensions have done, the magnitude only image reconstruction can meet the needs of most clinical applications.


In this paper, $\mathbb{R}$, $\mathbb{R}_+$ and $\mathbb{C}$ denote the sets of real, nonnegative real and complex numbers, respectively. $\langle \cdot,\cdot \rangle$ denotes the inner product of vectors.
$\preceq$ and $\succeq$ denote the elementwise operations of $\leq$ and $\geq$ on vectors, respectively. $\odot$ denotes the Hadamard or elementwise product of vectors. $| \cdot |$ takes
elementwise magnitude of vectors and $\angle$ unitizes elements of vectors, such that $\mathbf{v} = |\mathbf{v}|\odot \angle \mathbf{v}$ for a complex valued vector $\mathbf{v}$.
$\mathbf r = (x,y)$ and $\mathbf k=(k_x,k_y)$ denote the 2D coordinate systems of the spatial image and $k$-space domains, respectively.

\section{Theory}
\subsection{Formulation of undersampled $k$-space data}
Consider a pMRI scanner implemented with $L$ receiver coils. Let $h(\mathbf{r})\in \mathbb{C}$ be the 2D spatial MR image function and $s_i(\mathbf{r}) \in \mathbb{C}$, $i=1,\cdots, L$, be the 2D spatial sensitivity functions of the coils. The sensitivity encoded image functions of the coils are products of $h(\mathbf{r})$ and $s_i(\mathbf{r})$ written as
$$z_i(\mathbf{r})=h(\mathbf{r})s_i(\mathbf{r}) \in \mathbb{C}, \ i=1,\cdots, L.$$
The MRI scan creates the following $k$-space functions of the $L$ receiver coils
\begin{equation}\label{model1}
  g_i(\mathbf{k})=  \int \int z_i(\mathbf{r})e^{-j2\pi\langle\mathbf{k},\mathbf{r}\rangle}d\mathbf{r},\quad i=1,\cdots,L,
\end{equation}
which are the Fourier transforms of $z_i(\mathbf{r})=h(\mathbf{r})s_i(\mathbf{r})$. The $k$-space functions are undersampled and formulated into the following discrete data vectors
\begin{equation}\label{gi}
\mathbf g_i
=\mathbf{F} \mathbf{z}_i =
\mathbf{F} (\mathbf{s}_i\odot\mathbf{h}), \ \ i=1,2,\cdots,L,
\end{equation}
where $\mathbf{g_i}\in \mathbb{C}^{M}$, $M<N^2$, $i=1,\cdots, L$, are the undersampled $k$-space data vectors, $\mathbf{h}, \ \mathbf{s_i}, \ \mathbf{z_i} \in \mathbb{C}^{N^2}$ are the discretized vectors of $h(\mathbf{r})$, $s_i(\mathbf{r})$ and $z_i(\mathbf{r})$, respectively,  and the matrix $\mathbf{F} \in \mathbb{C}^{M \times N^2}$ is the corresponding undersampled partial 2D discrete Fourier transform (DFT) matrix operating on vectors.

For the purpose of magnitude image reconstruction, let ${\mathbf h}_m = |{\mathbf h}| \in \mathbb{R}_+^{N^2}$ be the magnitude of the image vector ${\mathbf h}$ and $\hat {\mathbf s}_i={\mathbf s}_i \odot \angle {\mathbf h} \in \mathbb{C}_+^{N^2}$,
$i=1, \cdots, L$. The undersampled $k$-space data vectors ${\mathbf g}_i$ can be written as
\begin{equation}\label{gi2}
\mathbf g_i
=\mathbf{F} \mathbf{z}_i =
\mathbf{F} (\hat {\mathbf{s}}_i\odot \mathbf{h}_m), \ \ i=1,2,\cdots,L.
\end{equation}
The objective of magnitude image reconstruction is to find a solution for $\mathbf{h}_m$ using given undersampled $k$-space data vectors ${\mathbf g}_i$, $i=1,\cdots, L$.
\subsection{The convex solution space}
Given the undersampled $k$-space data vectors ${\mathbf g}_i$, $i=1,\cdots, L$, in the form (\ref{gi}), to find a joint solution for the image vector ${\mathbf h}$ and the sensitivity functions ${\mathbf s}_i$ is in general a nonlinear and nonconvex problem. If only the magnitude of the image function is considered, it is possible to construct a convex solution space for the vector equation (\ref{gi2}), which can further lead to a formulation of a convex optimization problem for the magnitude image reconstruction. An observation of the convex solution space is introduced below.

\begin{figure}[!t]
\centerline
{\includegraphics[width=3in]{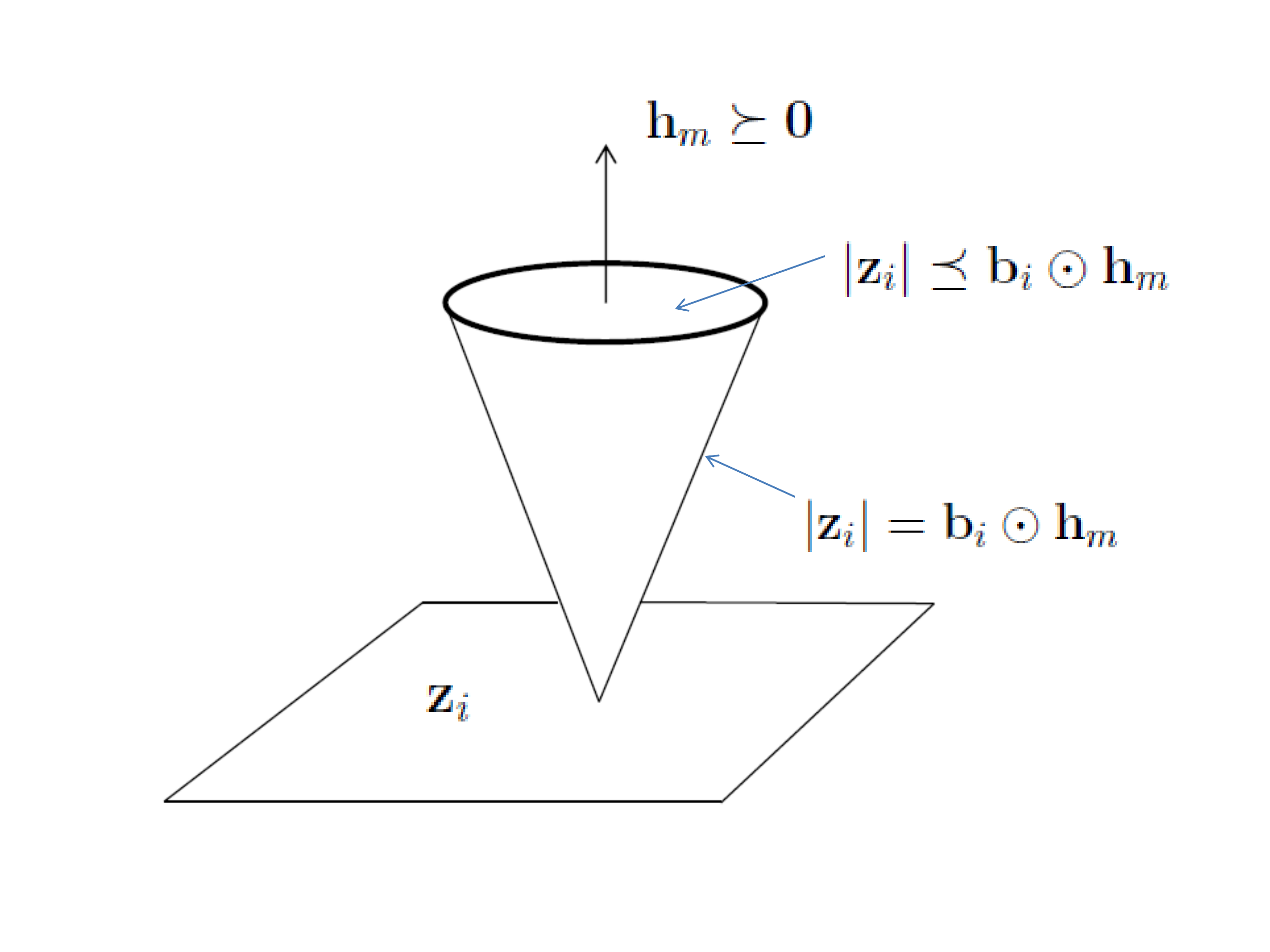}}
\caption{Convex solution space for ${\mathbf h}_m$ and ${\mathbf z}_i$. }
\label{cone}
\end{figure}

In each bilinear equation ${\mathbf z}_i=\hat {\mathbf{s}}_i \odot \mathbf{h}_m$ of the sensitivity encoded image functions, there are in general two independent variable vectors which, if known, can determine the third vector variable.
Since the sensitivity functions ${\mathbf s}_i$ have bounded magnitudes due to bounded inductances of the coils, there exist constant vectors ${\mathbf b}_i \in \mathbb{R}_+^{N^2}$ such that $|\hat {\mathbf s}_i| = |{\mathbf s}_i|\preceq {\mathbf b}_i$, $i=1, \cdots, L$. It follows that the magnitudes of ${\mathbf z}_i = \hat {\mathbf{s}}_i\odot \mathbf{h}_m$ are constrained by
\begin{equation}\label{bounds}
|{\mathbf z}_i| \preceq \mathbf{b}_i \odot \mathbf{h}_m, \ \ i=1,\cdots,L.
\end{equation}
If ${\mathbf h}_m$ and ${\mathbf z}_i$ are considered as independent variables, for some $i$, the inequality (\ref{bounds}) forms a cone shaped convex hull, which contains the true solutions of ${\mathbf h}_m$ and ${\mathbf z}_i$ for the vector equation (\ref{gi2}), if the
constant bound vector ${\mathbf b}_i$ is properly chosen. For the scalar case of ${\mathbf h}_m \in \mathbb{R}_+$ and ${\mathbf z}_i\in \mathbb{C}$, such a convex solution space for ${\mathbf h}_m$ and ${\mathbf z}_i$ is displayed in Fig. \ref{cone}, on top of the complex plane of ${\mathbf z}_i$. This convex solution space provides a basis for the convex optimization of the pMRI reconstruction problem and its extension to the high dimensional convex solution space is straightforward. It is, however, noted that
the convex solution space only exists for the positive valued magnitude image ${\mathbf h}_m$ but not for other real or complex valued image vectors.
\subsection{Convex optimization for image reconstruction}
With ${\mathbf h}_m$ and ${\mathbf z}_i$ as variables and with given ${\mathbf g}_i$ and the constraints (\ref{bounds}), $i=1,2,\cdots,L$, the $k$-space data equations (\ref{gi2}) are underdetermined and have in general infinite number of solutions for ${\mathbf h}_m$ and ${\mathbf z}_i$. Specifically, if $\{{\mathbf h}_{m0}, {\mathbf z}_{i0}, \ i=1,2,\cdots,L\}$ is a solution for (\ref{gi2}), $\{a{\mathbf h}_{m0}, {\mathbf z}_{i0}, \ i=1,2,\cdots,L\}$ is also a solution for any $a>1$. An effective approach to finding a meaningful solution for the
underdetermined equation (\ref{gi2}) is to apply regularized optimization to penalize ${\mathbf h}_m$ in terms of some normed function of ${\mathbf h}_m$.

Let ${\mathbf W}$ denotes a general wavelet transformation matrix and assume that the wavelet transformed image vector ${\mathbf {Wh}}_m$ is sparse. The $\ell_1$ regularized optimization
for the image reconstruction promotes the regularization variable and is formulated as
\begin{equation}\label{op1}
\begin{array}{l}
\min_{\mathbf{h}_m, \mathbf{z}_i, i=1,\cdots L } {1 \over 2}\sum_{i=1}^{L}\| {\mathbf{g}}_i - {\mathbf{F}} {\mathbf{z}_i}\|_2^2 + \alpha \|{\mathbf {Wh}}_m\|_1 \\ \\
\textrm{subject to:} \ \ {\mathbf {h}}_m \succeq {\mathbf 0}, \ \
|{\mathbf z}_i| \preceq \mathbf{b}_i \odot \mathbf{h}_m, \ i=1,\cdots,L,
\end{array}
\end{equation}
where $\alpha\geq 0$ is a regularization parameter. The $\ell_1$ regularized optimization problem in (\ref{op1}) can alternatively be formulated as
\begin{equation}\label{op2}
\begin{array}{l}
\min_{\mathbf{h}_m, \mathbf{z}_i, \mathbf{w}_i, \ i=1,\cdots L } \  {1 \over 2}\sum_{i=1}^{L}\| {\mathbf{g}}_i - {\mathbf{F}} {\mathbf{z}_i}\|_2^2\\ \\
 + \alpha \|{\mathbf {Wh}}_m\|_1
 + \beta \sum_{i=1}^{L}\| |{\mathbf z}_i| - \mathbf{b}_i \odot \mathbf{h}_m + \mathbf{q}_i\|_2^2 \\ \\
\textrm{subject to:} \ \ {\mathbf {h}}_m \succeq {\mathbf 0}, \ \
{\mathbf {q}}_i \succeq {\mathbf 0}, \ \ i=1,\cdots,L,
\end{array}
\end{equation}
where $\beta\geq 0$ is a regularization parameter.

The above formulated regularized optimization problem is convex and can yield a globally optimal and unique solution for the magnitude image ${\mathbf h}_m$ and the sensitivity encoded image functions ${\mathbf z}_i$, $i=1,\cdots, L$. It is noted that the $\ell_1$ regularized optimization promotes the sparsity of the regularization variable. It is possible to select other regularization terms of ${\mathbf h}_m$ and ${\mathbf z}_i$ to meet reconstruction specifications when needed.


\section{Methods}
\subsection{$k$-space data sets}
The proposed convex optimization method is applied to two sets of \emph{in-vivo} MR data for reconstruction and performance evaluation. The first brain data set was acquired on a 3T scanner (GE Healthcare, Waukesha, WI) using an $8$-channel head channel (Invivo, Gainesville, FL) and a 2D $T1$-weighted spin echo protocol with axial plane, $TE/TR = 11/700$ ms, FOV=$22 cm^2$, $10$ slices and $256 \times 256$ matrix size. The second spine data set was acquired using an $8$-channel cervical-thoracic-lumbar spine array and a fast spoiled gradient-echo sequence with $TR/TE = 300/12$ ms, RBW $= 62.5$ kHz, $256 \times 256$, Tip angle $= 15^{\circ}$ and FOV = $32 \times 32 cm^2$. The fully sampled $k$-space data sets were in the cartesian coordinate system and were manually undersampled by uniform sampling with additional auto-calibration signal (USACS) lines in the phase encoding direction.
\subsection{Computing Set Ups}
The computation of the proposed optimization problem (\ref{op2}) is implemented with the split Bregman method for iterative $l_1$ regularized optimization in \cite{goldstein_2009}.
The regularization parameters are empirically chosen and a tolerance value of $10^{-5}$ is selected for each step of iteration. The haar wavelet matrix is adopted for $\mathbf W$. The initial values of ${\mathbf h}_m$ and ${\mathbf z}_i$, $i=1,\cdots,L$, are randomly chosen in each iterative reconstruction. And the algorithm is programmed with Matlab (Math-Works, Natick, MA, USA).

To evaluate the reconstruction performance, the reconstructed images, denoted by $\mathbf h^o$, are compared with the sum of square (SOS) image, which is reconstructed using the fully sampled data and denoted as $\mathbf h_{SOS}$. The normalized mean square error (NMSE) of $\mathbf h^o$ is defined as
$$e_{NMSE}={{\|\mathbf h^o-\mathbf h_{SOS}\|_2^2}\over {\|\mathbf h_{SOS}\|_2^2}}.$$
The NMSE values of reconstructed images by the proposed algorithm are computed and compared with that by other recently developed methods
for the in-vivo brain data sets under the same data reduction conditions.
\begin{figure}[!htb]
    \centerline
   { { \includegraphics[scale=0.45]{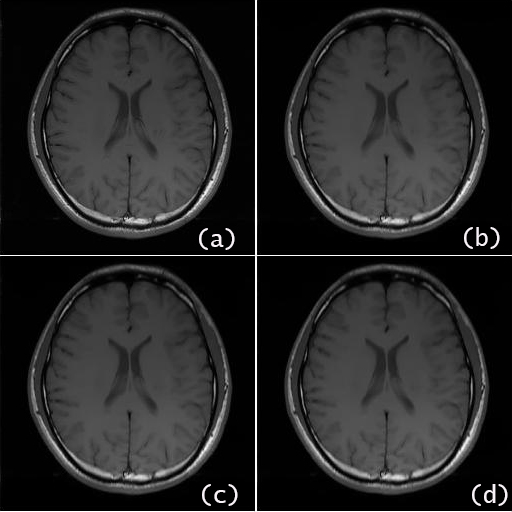}}}
\caption{Reconstructed brain images by our method, (a) SOS image as reference, (b) Reconstructed image with $f_{nom}=4$, (c)  Reconstructed image with $f_{nom}=8$, (d)  Reconstructed image with $f_{nom}=12$.}\label{brain}
\end{figure}
\begin{figure}[!htb]
    \centerline
   { { \includegraphics[scale=0.45]{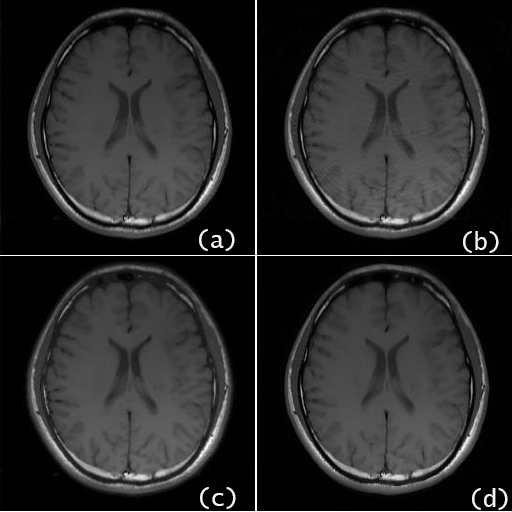}}}
\caption{Comparison of brain image reconstructions with $f_{nom}=8$, (a) by our proposed method, (b) by $L_1$ CG SPIRiT, (c)  by Sparse BLIP, (d)  by IRGN TGV.}\label{comp}
\end{figure}
\begin{figure}[!htb]
    \centerline
   { { \includegraphics[scale=0.45]{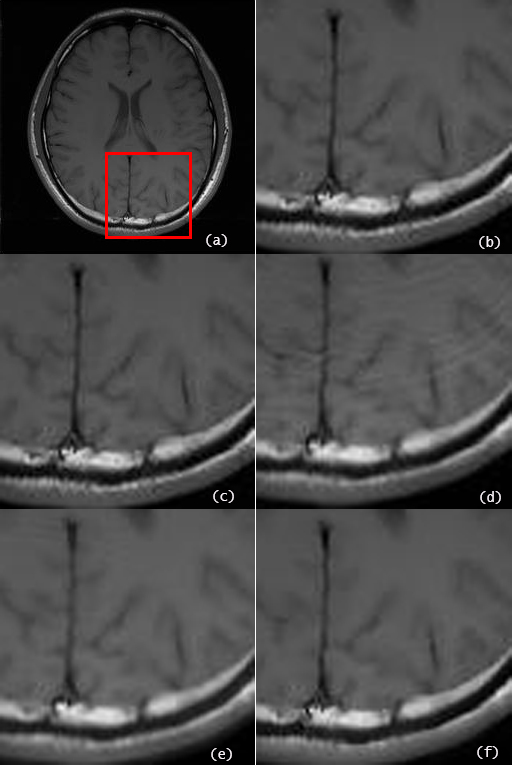}}}
\caption{Comparison of a zoomed portion of brain image reconstructions with $f_{nom}=8$, (a) SOS image as reference, (b) Zoomed portion of SOS, (c) by our proposed method, (d)  by SPIRiT, (e) by Sparse BLIP and (f) by IRGN-TGV.}\label{zcomp}
\end{figure}
\begin{figure}[!htb]
    \centerline
   { { \includegraphics[scale=0.45]{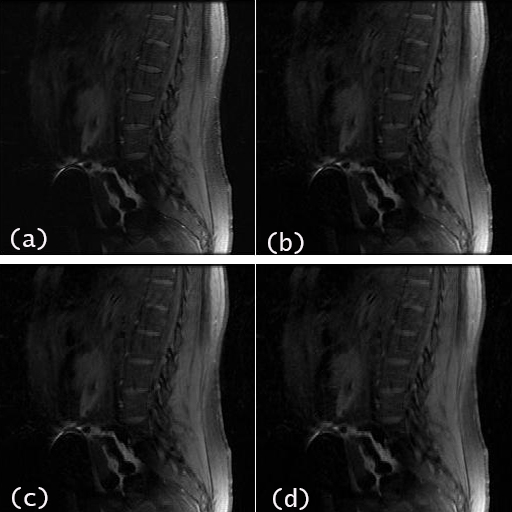}}}
\caption{Reconstructed spine images by our proposed method, (a) SOS image as reference, (b) Reconstructed image with $f_{nom}=4$, (c)  Reconstructed image with $f_{nom}=8$, (d)  Reconstructed image with $f_{nom}=12$.}\label{spine}
\end{figure}

\begin{table}[!t]
\renewcommand{\arraystretch}{1.3}
\caption{NMSE's of different algorithms}
\label{trf}
\centering
\begin{tabular}{|l|r|r|r|r|}
\hline
\bfseries Method  & \bfseries  IRGN-TGV & \bfseries Sparse-BLIP & \bfseries  SPIRiT  & \bfseries Our Method\\
\hline\hline
 $f_{nom}=4$   & 0.0042   & 0.0040   & 0.0066     & 0.0038\\
 $f_{nom}=8$     & 0.0056   & 0.0048   & 0.0102     & 0.0045\\
$f_{nom}=12$     & 0.0086   & 0.0072   & 0.0148     & 0.0063\\
\hline
\end{tabular}
\end{table}

$\sqrt{2}$
\section{Results}
For the brain data set, the reconstructed images by the proposed algorithm using undersampled $k$-space data of nominal reduction factor, $f_{nom}=4 \ (net reduction factor, f_{net}=2.66)$ and higher $f_{nom}=8\ (f_{net}=4.00)$ and $f_{nom}=12\ (f_{net}=4.83)$ are shown in
Fig.\ref{brain} (b), (c) and (d), respectively, in comparison with the SOS image using the full $k$ space data in Fig.\ref{brain} (a).
For the brain date set of $f_{nom}=8$, Fig.\ref{comp} presents the reconstructed image by the proposed method in comparison with that by other recent algorithms, which are IRGN-TGV \cite{knoll_2012}, SPIRiT \cite{lustig_2010} and Sparse-BLIP \cite{she_2012}. To provide more detailed reconstruction information for comparison, a selected area of each reconstructed image by different algorithms is zoomed and displayed in Fig. \ref{zcomp}.

The NMSEs of the reconstructed brain images of different algorithms are listed in Table \ref{trf} for $f_{nom}=4, \ 8, \ 12$, respectively. The results clearly show that the proposed convex optimization method outperforms other methods in terms of the reconstruction accuracy at high undersampling rates as well as reduction of artifacts.

For the spine data set,
the reconstructed image by SOS is given in Fig.\ref{spine} (a) followed by that reconstructed using the proposed method Fig. \ref{spine} (b), for nominal undersampling rate $f_{nom}=4 \ (f_{net}=2.66)$. The NMSE
of this reconstructed image is $e_{NMSE}=0.0036$. For higher reduction rates of $f_{nom}=8 \ (f_{net}=4.00)$ and $f_{nom}=12 \ (f_{net}=4.83)$, the corresponding reconstructed images by the proposed algorithm are shown in Fig.\ref{spine} (c) and \ref{spine} (d) with NMSE values $0.0045$ and $0.0061$, respectively.

With randomly selected different initial values for the magnitude image and sensitivity encoded images for the proposed convex regularization algorithm, the reconstructed brain and spine images with different undersampling rates always converge to the same solutions. These results demonstrated the globally unique and optimal solution of the proposed convex optimization. In contrast, simulations of other nonconvex algorithms resulted in significantly different reconstruction solution under differently initial conditions.

\section{Conclusion}
The pMRI reconstruction based on undersampled $k$-space data by optimization methods is a nonlinear and nonconvex problem. The existing optimization methods for jointly computing the image and sensitivity functions can only provide local but not global solutions. The computing of such a nonlinear and nonconvex problem involve complicated procedures and iterations. It is shown in this paper that, if only the magnitude image reconstruction is considered, there is a convex solution space for the magnitude image and sensitivity encoded image functions. This enables formulation of a regularized convex optimization problem for the magnitude image reconstruction and the solution of this problem is globally unique and optimal. Extensive simulations of in-vivo data sets have been carried out and the superior performance of the proposed convex optimization for pMRI reconstruction has been shown.

\bibliographystyle{IEEEtran}
\bibliography{reffilef}

\begin{thebibliography}{10}
\providecommand{\url}[1]{#1}
\csname url@samestyle\endcsname
\providecommand{\newblock}{\relax}
\providecommand{\bibinfo}[2]{#2}
\providecommand{\BIBentrySTDinterwordspacing}{\spaceskip=0pt\relax}
\providecommand{\BIBentryALTinterwordstretchfactor}{4}
\providecommand{\BIBentryALTinterwordspacing}{\spaceskip=\fontdimen2\font plus
\BIBentryALTinterwordstretchfactor\fontdimen3\font minus
  \fontdimen4\font\relax}
\providecommand{\BIBforeignlanguage}[2]{{%
\expandafter\ifx\csname l@#1\endcsname\relax
\typeout{** WARNING: IEEEtran.bst: No hyphenation pattern has been}%
\typeout{** loaded for the language `#1'. Using the pattern for}%
\typeout{** the default language instead.}%
\else
\language=\csname l@#1\endcsname
\fi
#2}}
\providecommand{\BIBdecl}{\relax}
\BIBdecl

\bibitem{roemer_1990}
P.~B. Roemer, W.~A. Edelstein, C.~E. Hayes, S.~P. Souza, and O.~M. Mueller,
  ``{The NMR Phased Array},'' \emph{{Magnetic Resonance in Medicine}},
  vol.~{16}, pp. {192--225}, {1990}.

\bibitem{pruessmann_1999}
K.~P. Pruessmann, M.~Weiger, M.~B. Scheidegger, and P.~Boesiger, ``{SENSE:
  Sensitivity encoding for fast MRI},'' \emph{{Magnetic Resonance in
  Medicine}}, vol.~{42}, pp. {952--962}, {1999}.

\bibitem{griswold_2002}
M.~Griswold, P.~Jakob, R.~Heidemann, M.~Nittka, V.~Jellus, J.~Wang, B.~Kiefer,
  and A.~Haase, ``{Generalized Autocalibrating Partially Parallel Acquisitions
  (GRAPPA)},'' \emph{{Magnetic Resonance in Medicine}}, vol.~{47}, pp.
  {1202--1210}, {2002}.

\bibitem{lustig_2010}
M.~Lustig and J.~M. Pauly, ``{SPIRiT: Iterative Self-consistent Parallel
  Imaging Reconstruction From Arbitrary k-Space},'' \emph{{Magnetic Resonance
  in Medicine}}, vol.~{64}, pp. {457--471}, {2010}.

\bibitem{park_2012}
S.~Park and J.~Park, ``{Adaptive self-calibrating iterative GRAPPA
  reconstruction},'' \emph{{Magnetic Resonance in Medicine}}, vol.~{67}, pp.
  {1721--1729}, {2012}.

\bibitem{uecker_2008}
M.~Uecker, T.~Hohage, K.~Block, and J.~Frahm, ``{Image reconstruction by
  regularized nonlinear inversion - Joint estimation of coil sensitivities and
  image content},'' \emph{{Magnetic Resonance in Medicine}}, vol.~{60}, pp.
  {674--682}, {2008}.

\bibitem{knoll_2012}
F.~Knoll, C.~Clason, K.~Bredies, M.~Uecker, and R.~Stollberger, ``{Parallel
  imaging with nonlinear reconstruction using variational penalties},''
  \emph{{Magnetic Resonance in Medicine}}, vol.~{67}, pp. {34--41}, {2012}.

\bibitem{she_2012}
H.~She, R.~Chen, D.~Liang, E.~DiBella, and L.~Ying, ``{Simultaneous image
  reconstruction and sensitivity estimation in parallel MRI using blind
  compressed sensing},'' {2012}, pp. { 876 -- 879 }, {\em 9th IEEE
  International Symposium on Biomedical Imaging (ISBI) }.

\bibitem{cishen_2013}
C.~Zhang and I.~Baqee, ``{Parallel magnetic resonance imaging reconstruction by
  convex optimization },'' {2013}, pp. {473--478}, {2013 Third International
  Conference on Innovative Computing Technology (INTECH) }.

\bibitem{goldstein_2009}
T.~Goldstein and S.~Osher, ``{The split Bregman method for $l_1$ regularized
  problems},'' \emph{{SIAM Journal of Imaging Science}}, vol.~{2}, pp.
  {323--343}, {2009}.

\end{thebibliography}

\end{document}